# Θ-point behavior of diluted polymer solutions: Can one observe the universal logarithmic corrections predicted by field theory?

Johannes Hager and Lothar Schäfer
*Fachbereich Physik, Universität Essen, 45117 Essen, Germany*

In recent large-scale Monte Carlo simulations of various models of Θ-point polymers in three dimensions Grassberger and Hegger found logarithmic corrections to mean field theory with amplitudes much larger than the universal amplitudes of the leading logarithmic corrections calculated by Duplantier in the framework of tricritical $O(n)$ field theory. To resolve this issue we calculate the universal subleading correction of field theory, which turns out to be of the same order of magnitude as the leading correction for all chain lengths available in present day simulations. Borel resummation of the renormalization group flow equations also shows the presence of such large corrections. This suggests that the published simulations did not reach the asymptotic regime. To further support this view, we present results of Monte Carlo simulations on a Domb-Joyce-like model of weakly interacting random walks. Again the results cannot be explained by keeping only the leading corrections, but are in fair accordance with our full theoretical result. The corrections found for the Domb-Joyce model are much smaller than those for other models, which clearly shows that the effective corrections of are not yet in the asymptotic regime. Altogether, our findings show that the existing simulations of Θ polymers are compatible with tricritical field theory, since the crossover to the asymptotic regime is very slow. Similar results were found earlier for self-avoiding walks at their upper critical dimension $d_c=4$. [S1063-651X(99)15108-0]

## I. INTRODUCTION

The behavior of long polymers in good solvents can be described quantitatively with the ingredients of chain connectedness and excluded volume interaction. De Gennes' mapping on the $O(n)$-symmetric $\phi^4$ field theory in the limit $n=0$ provided scaling laws like $R_g \sim N^{\nu}$ for the radius of gyration or $Z \sim e^{\mu^* N} N^{\gamma-1}$ for the partition sum with anomalous exponents $\nu=0.588$ and $\gamma=1.157$ (in $d=3$). These asymptotic laws hold at the critical point reached for chain length $N\to\infty$ in the excluded volume region, where the effective interaction among the polymer segments is repulsive. The coil globule Θ transition at some temperature $T_\Theta$ separates this regime from another one, where the segments attract each other and the polymer forms a compact globule with $R_g \sim N^{1/d}$. The Θ transition is generally believed to be a second-order transition in the sense that the segment density within an infinitely large globule increases continuously from zero with decreasing temperatures $T<T_\Theta$. Again de Gennes [1,2] proposed a mapping on $O(n)$-symmetric $\phi^6$ field theory in the limit $n=0$ thereby identifying the Θ point as a tricritical point for $N\to\infty$ with mean field exponents $\nu=1/2$, $\gamma=1$ at the upper critical dimension $d_c=3$. Duplantier [3,4] calculated the leading universal logarithmic corrections to mean field theory for several polymer observables. A measurement of these corrections is the ultimate test for the tricritical approach to Θ collapse. Unfortunately, the precision of present experiments is not high enough for that purpose. So up to now only computer simulations have a chance to detect the logarithmic corrections. Indeed large-scale simulations of long lattice walks [5,6] found corrections, which can be interpreted as being logarithmic, but which show a much larger amplitude than predicted. In contrast, simulations of off-lattice models [7,8] claimed to find agreement with [3,4] using the lowest-order formula

$$\overline{w}_R=\frac{\overline{w}_0}{1+\frac{22}{15}\,\overline{w}_0\ln(N s_{m^2})} \tag{1}$$

for the $N$ dependence of the renormalized coupling constant $\overline{w}_R$. However, this interpretation is somewhat ambiguous since (1), although it may be a reasonable interpolation between the starting value $\overline{w}_0$ and the asymptotic behavior $\sim 15/(22\ln(N))$ of the renormalized coupling for large $N$, it does not have the proper expansion in terms of the chain length. Especially it does not contain the subleading universal correction proportional to $\ln[\ln(N)]/\ln^2(N)$ calculated in the sequel. For a proper use of Eq. (1) one has to assure that the chain lengths under consideration are large enough to suppress the subleading logarithmic corrections, which we will find to be an impossible task for simulations as well as for experiments.

It should be noted that the tricritical framework is not the only scenario for the collapsing transition of polymer chains. Other scenarios have been proposed, in which this transition is of first order and leads to an intermediate phase where the polymer configuration shows a branched structure [9]. The collapse can also lead to a frozen phase [10,11], where the segment directions are ordered. Which scenario is realized depends on the details of the microstructure of the chain. The aim of the present paper is to clarify whether the different findings of computer simulations for the standard models can be explained within the tricritical framework.

The paper is organized as follows: In Sec. II we briefly describe Edwards model of a continuous polymer chain at the $\Theta$ point and its relation to the tricritical $\phi^6$ theory, and we present results for the renormalization factors calculated in Appendix A. In Sec. III we evaluate the renormalization group (RG) flow functions. In the polymer limit $n=0$ we calculate the subleading logarithmic corrections to mean field theory due to the RG flow and also give resummed forms of the flow functions. In Sec. IV we review first-order results for polymer scaling functions. Section V describes a modified Domb-Joyce model for $\Theta$-point polymers and the Monte Carlo algorithm used for the simulations. In Sec. VI we compare our simulation results to the field-theoretic predictions of Secs. III and IV. In Sec. VII we give our conclusion and suggest further developments.

## II. EDWARDS MODEL, FIELD THEORY, AND RENORMALIZATION

We use the Edwards model, where the configuration of a polymer chain is described as a continuous curve $\mathbf{r}(s)$ in $d$-dimensional space. The parameter $s$ is taken to have dimensions of (length)$^2$, and the total ‘‘length’’ of the chain equals the ‘‘Gaussian surface’’ $S$. The interaction energy of a single chain formally is written as

$$E(\mathbf{r})=\frac{1}{4}\int_0^S ds\left(\frac{d\mathbf{r}(s)}{ds}\right)^2+\frac{u_0}{4!}\int_0^S ds_1\int_0^S ds_2\,\delta^d(\mathbf{r}(s_1)-\mathbf{r}(s_2))+\frac{w_0}{6!}\int_0^S ds_1\int_0^S ds_2\int_0^S ds_3\,\delta^d(\mathbf{r}(s_1)-\mathbf{r}(s_2))\times\delta^d(\mathbf{r}(s_2)-\mathbf{r}(s_3)),\tag{2}$$

where the first term describes the connectivity of the chain, and where we include local two-body and three-body interactions of strengths $u_0$ and $w_0$, respectively. In the form written, the model is singular and must be regularized by introducing a cutoff $\Delta s=l_0^2$, which gives the minimal distance along the chain of two interacting segments. With this regularization the model essentially is equivalent to a discrete chain of $N=S/l_0^2$ segments of average length $\sqrt{2d}l_0$. At the $\Theta$ point the two-body coupling $u_0$ is attractive. On macroscopic scales it has to compensate the effects of the repulsive three-body coupling $w_0$: $u_0(\Theta)=u_\Theta(w_0,l_0)<0$. For $u_e=u_0-u_\Theta>0$ we are in the excluded volume regime of swollen coils, and for $u_e<0$ the chain collapses. Right at $u_e=0$ the chains for $d\geqslant 3$ show unperturbed random walk behavior in the asymptotic limit $S\to\infty$. This is the ‘‘tricritical’’ scenario.

As is well known [1,2], the Edwards model can be mapped on a local field theory built upon an $n$-component field $\phi_\alpha$, $\alpha=1,\ldots,n$ (Landau-Ginzburg-Wilson model). The resulting Hamiltonian is $O(n)$ symmetric and reads

$$E(\phi)=\int d^dx\,\frac{1}{2}[(\nabla\phi)^2+m_0^2\phi^2]+\frac{u_0}{4!}(\phi^2)^2+\frac{w_0}{6!}(\phi^2)^3.\tag{3}$$

The mapping proceeds via Laplace transformation replacing the Gaussian surface $S$ by the conjugate ‘‘mass’’ parameter $m_0^2$. In the field theory the formal limit $n\to 0$ has to be taken, which exists order by order in perturbation theory. The thermodynamic properties of the field-theoretic model are inherent in the partition function $Z$ which is defined as a functional integral over the field $\phi$:

$$Z=\int D[\phi(\mathbf{x})]e^{-E(\phi)}.\tag{4}$$

Local observables are the expectation values or Greens functions:

$$\langle\phi_{\alpha_1}(\mathbf{x}_1)\cdots\phi_{\alpha_{2M}}(\mathbf{x}_{2M})\rangle=G^{(2M)}_{\alpha_1,\ldots,\alpha_{2M}}(\mathbf{x}_1,\ldots,\mathbf{x}_{2M})=\frac{1}{Z}\int D[\phi(\mathbf{x})]\phi_{\alpha_1}(\mathbf{x}_1)\cdots\phi_{\alpha_{2M}}(\mathbf{x}_{2M})e^{-E(\phi)}.\tag{5}$$

The limit of infinitely long chains corresponds to a critical point of the field theory, identified by a critical value $m_{0c}^2$ of the mass. At $m_0^2=m_{0c}^2$ the theory becomes invariant under a change of the length scale. However, like the Edwards model the field theory *a priori* has to be regularized by a small distance cutoff $\sim l_0$, and to extract the scale-invariant properties $l_0$ has to be eliminated: $l_0\to 0$. Order for order in perturbation theory this can be achieved by the standard method of renormalization. In the sequel we restrict ourselves to the $\Theta$ point $u_e=u_0-u_\Theta=0$, and we consider the theory in the neighborhood of the physical dimension $d=3$. We introduce renormalized fields and renormalized parameters via the formal relations

$$\phi=Z_\phi^{1/2}l_R^{1-d/2}\phi_R,\tag{6}$$

$$\frac{w_0}{2^5\pi^2}=Z_w l_R^{2(d-3)}\overline{w}_R=\frac{Z_6}{Z_\phi^3}l_R^{2(d-3)}\overline{w}_R,\tag{7}$$

$$m_0^2-m_{0c}^2=Z_{m^2}l_R^{-2}m_R^2=\frac{Z_2}{Z_\phi}l_R^{-2}m_R^2.\tag{8}$$

Here $l_R$ is an arbitrary length scale introduced to make all renormalized quantities dimensionless and a geometric factor $1/(2^5\pi^2)$ has been absorbed in the renormalized coupling $\overline{w}_R$. The renormalization factors $Z_\phi$, $Z_w$, and $Z_{m^2}$ are functions of $\overline{w}_R$ that will be specified below. With these formal substitutions the Greens functions (5) take the form

$$G_c^{(2M)}(\mathbf{r}_1,\ldots,\mathbf{r}_{2M},m_0^2,w_0,l_0)=Z_\phi^M l_R^{M(2-d)}\overline{G}^{(2M)}_{c,R}\left(\frac{\mathbf{r}_1}{l_R},\ldots,\frac{\mathbf{r}_{2M}}{l_R},m_R^2,\overline{w}_R\right),\tag{9}$$

where $\overline{G}^{(2M)}_{c,R}$ is the dimensionless renormalized Greens function. In renormalization theory one proves that the renormalization factors can be constructed as formal power series in $\overline{w}_R$, with coefficients that absorb the leading cutoff dependence of the theory. The resulting renormalized Greens functions for $d\leqslant 3$ are finite in the limit $l_0\to 0$ order for order in their expansion in powers of $\overline{w}_R$.

Explicit low-order expressions for the $Z$ factors can be found in the literature [12]. To extract the subleading correc-tions for the behavior of the theory at the tricritical point we have pushed these calculations one order further, which amounts to four-loop or six-loop calculations. $Z_6$ is calculated including $O(\overline{w}_R^2)$, $Z_\phi$ and $Z_2$ including $O(\overline{w}_R^3)$. Technically we used the method of dimensional regularization which exploits the fact that for $d<3$ even the unrenormalized theory is finite in the limit $l_0 \to 0$. The leading microstructure information then is contained in pole type singularities found for

$$\epsilon_3 = 3 - d \to 0. \tag{10}$$

In the minimally subtracted renormalization scheme the $Z$ factors have to absorb these pole terms. Details of the calculations are given in Appendix A. We find

$$Z_\phi = 1 - \frac{1}{\epsilon_3}\frac{(n+2)(n+4)}{10800}\overline{w}_R^2 + \left(\frac{2}{3\epsilon_3} - \frac{1}{\epsilon_3^2}\right) \times \frac{(n+2)(n+4)(3n+22)}{243000}\overline{w}_R^3 + O(\overline{w}_R^4), \tag{11}$$

$$Z_2 = 1 + \frac{1}{\epsilon_3}\frac{(n+2)(n+4)}{720}\overline{w}_R^2 - \left(\frac{11}{\epsilon_3} - \frac{5}{2\epsilon_3^2}\right) \times \frac{(n+2)(n+4)(3n+22)}{40500}\overline{w}_R^3 - \frac{1}{\epsilon_3} \times \frac{\pi^2(n^4+24n^3+172n^2+480n+448)}{324000}\overline{w}_R^3 + O(\overline{w}_R^4), \tag{12}$$

$$Z_6 = 1 + \frac{1}{\epsilon_3}\frac{(3n+22)}{15}\overline{w}_R + \frac{1}{\epsilon_3^2}\frac{9n^2+132n+484}{225}\overline{w}_R^2 - \frac{1}{\epsilon_3}\left(\frac{\pi^2(n^3+34n^2+620n+2720)}{7200} + \frac{71n^2+1146n+4408}{1200}\right)\overline{w}_R^2 + O(\overline{w}_R^3). \tag{13}$$

## III. RENORMALIZATION GROUP FLOW

To exploit the consequences of scale invariance at the tricritical point we differentiate Eq. (9) with respect to the arbitrary renormalized length scale $l_R$. With the standard definitions

$$W(\overline{w}_R) = -l_R \frac{\partial \overline{w}_R}{\partial l_R}, \tag{14}$$

$$\eta(\overline{w}_R) = -l_R \frac{\partial}{\partial l_R}\ln Z_\phi, \tag{15}$$

$$2 - \frac{1}{\nu(\overline{w}_R)} = \gamma_{m^2}(\overline{w}_R) = l_R \frac{\partial}{\partial l_R}\ln Z_{m^2}, \tag{16}$$

as well as the relation

$$l_R \frac{\partial m_R^2}{\partial l_R} = \frac{m_R^2}{\nu(\overline{w}_R)}, \tag{17}$$

and the replacement $\mathbf{r}_i / l_R \to \mathbf{r}_i$ this leads to the RG equation

$$\left\{ -\sum_{i=1}^{2M} \mathbf{r}_i \cdot \nabla_{\mathbf{r}_i} - W(\overline{w}_R)\frac{\partial}{\partial \overline{w}_R} + \frac{m_R^2}{\nu(\overline{w}_R)}\frac{\partial}{\partial m_R^2} - M[d-2+\eta(\overline{w}_R)] \right\} \overline{G}_{c,R}^{(2M)} = 0. \tag{18}$$

For $m_R^2 = 0$ and $W(\overline{w}_R) = 0$ this is precisely the Ward identity of the dilatation group. Therefore, the massless theory at a fixed point $\overline{w}_R(l_R) = w^*$, where $W(\overline{w}_R^*) = 0$, is scale invariant. Using Eqs. (7), (11), and (13) we find for the Wilson function (14)

$$W(\overline{w}_R) = \frac{-2\epsilon_3 \overline{w}_R}{1 + \overline{w}_R \dfrac{\partial \ln Z_w}{\partial \overline{w}_R}} = -2\epsilon_3 \overline{w}_R + \frac{6n+44}{15}\overline{w}_R^2 - \left(\frac{\pi^2(n^3+34n^2+620n+2720)}{1800} + \frac{53n^2+858n+3304)}{225}\right)\overline{w}_R^3 + O(\overline{w}_R^4). \tag{19}$$

Similarly we find for the exponent functions

$$\eta(\overline{w}_R) = W(\overline{w}_R)\frac{\partial \ln Z_\phi}{\partial \overline{w}_R} = \frac{(n+2)(n+4)}{2700}\overline{w}_R^2 - \frac{(n+2)(n+4)(3n+22)}{60750}\overline{w}_R^3 + O(\overline{w}_R^4), \tag{20}$$

$$\nu(\overline{w}_R) = \frac{1}{2 + W(\overline{w}_R)\dfrac{\partial \ln Z_{m^2}}{\partial \overline{w}_R}} = \frac{1}{2} + \frac{(n+2)(n+4)}{675}\overline{w}_R^2 - \left(\frac{\pi^2(n^4+24n^3+172n^2+480n+448)}{216000} + \frac{3n^3+40n^2+156n+176}{2430}\right)\overline{w}_R^3 + O(\overline{w}_R^4). \tag{21}$$

In the $\epsilon_3$ expansion we find for the nontrivial infrared-stable fixed point emerging for $d<3$ the result

$$\overline{w}_R^* = \frac{15\epsilon_3}{3n+22} + \frac{15}{16}\,\frac{\pi^2(n^3+34n^2+620n+2720)+424n^2+6864n+26432}{(3n+22)^3}\,\epsilon_3^2 + O(\epsilon_3^3). \tag{22}$$

The fixed point values of the exponents take the form

$$\eta(\overline{w}_R^*) = \frac{1}{12}\,\frac{(n+2)(n+4)}{(3n+22)^2}\,\epsilon_3^2\left(1+\frac{\pi^2(3n^3+102n^2+1860n+8160)+1128n^2+18480n+71552}{24(9n^2+132n+484)}\,\epsilon_3\right)+O(\epsilon_3^4), \tag{23}$$

$$\nu(\overline{w}_R^*) = \frac{1}{2}+\frac{1}{3}\,\frac{(n+2)(n+4)}{(3n+22)^2}\,\epsilon_3^2 + \frac{\pi^2(-3n^5+114n^4+10572n^3+114072n^2+403584n+433536)}{576(3n+22)^2(9n^2+132n+484)}\,\epsilon_3^3$$
$$+\frac{(2976n^4+76992n^3+625792n^2+1956096n+1977344)}{576(3n+22)^2(9n^2+132n+484)}\,\epsilon_3^3+O(\epsilon_3^4). \tag{24}$$

The result for $\eta$ coincides with that calculated in [13] using another renormalization scheme with cutoff regularization and normalization conditions for finite mass. The $\epsilon_3^3$ term for $\nu$ is a new result.

### A. Resummation

The series expansions of $W$, $\eta$, and $\nu$ are known to be alternating asymptotic series in the renormalized coupling $\overline{w}_R$ [14,15]. Figure 1 displays the first-order (uppermost curve) and second-order (lowest curve) result for $W(\overline{w}_R)$ at $d=3$ in the polymer limit $n=0$. While the first-order result shows asymptotically free behavior in the infrared limit with $\overline{w}_R=0$ being the only fixed point, the second-order result displays an additional infrared-unstable fixed point at a finite coupling $\overline{w}_R$, which is believed to be a spurious result. To eliminate it we have to resum the series.

The leading behavior of the series coefficients $\beta_k$ of $W(\overline{w}_R)$ in a $\phi^{2r}$ theory at high order $k$ of perturbation theory is of the form

$$\beta_k=[k(r-1)]!\,a^k k^b c\left[1+O\left(\frac{1}{k}\right)\right], \tag{25}$$

as one finds via a semiclassical calculation for small negative coupling [14,15]. For a $\phi^6$ theory the parameters $a$ and $b$ in our convention are [15]

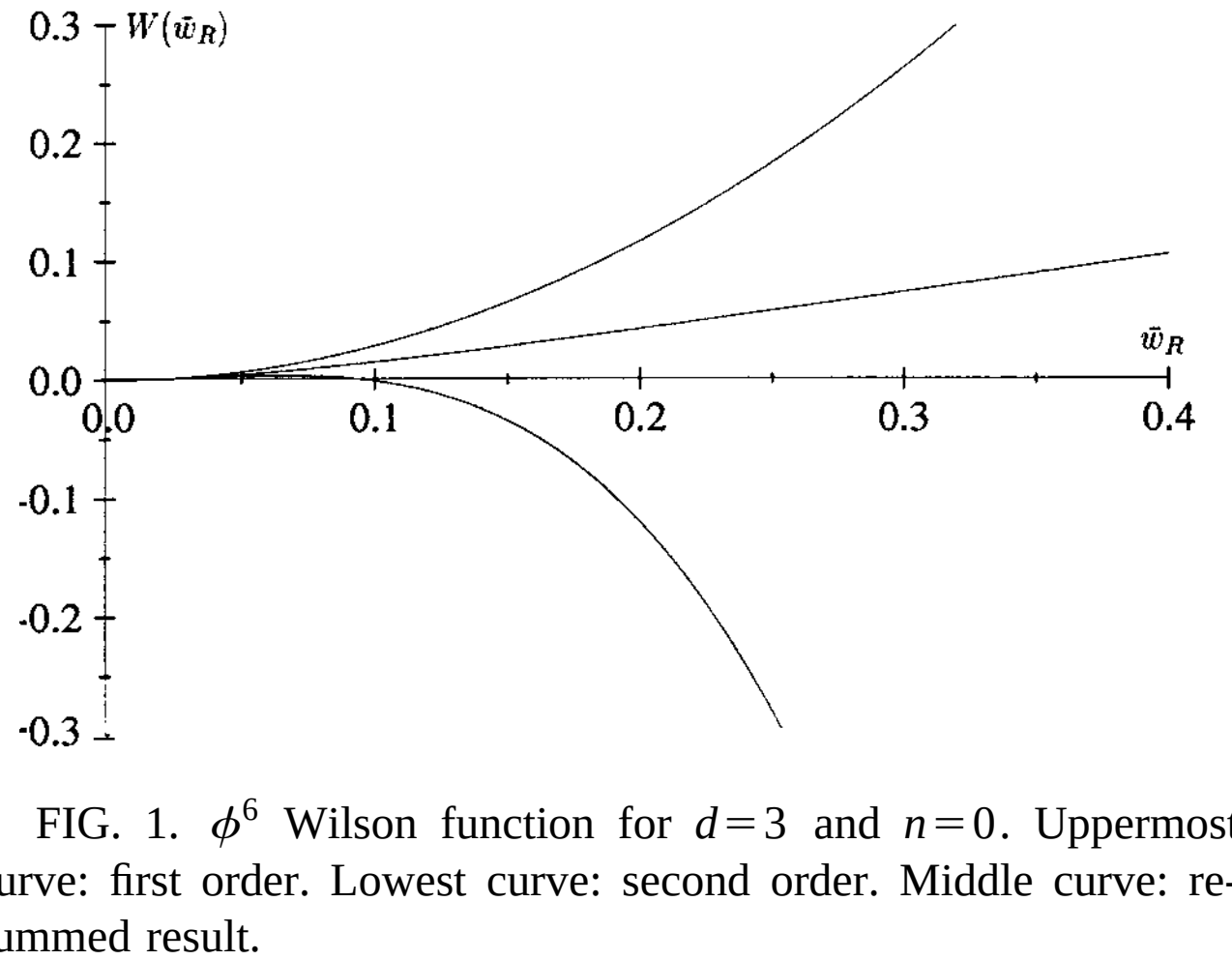

FIG. 1. $\phi^6$ Wilson function for $d=3$ and $n=0$. Uppermost curve: first order. Lowest curve: second order. Middle curve: resummed result.

$$a=-\frac{64}{45\pi^2},\qquad b=\frac{7+n}{2}. \tag{26}$$

With these parameters at hand we perform a standard Borel resummation procedure as described in [16,17]. Since our interest is in corrections due to finite chain lengths $N$ and therefore finite renormalized coupling $\overline{w}_R$ we resummed the Wilson function in $d=3$ for couplings in an interval $\overline{w}_R \in [0,1]$ in the polymer limit $n=0$. The resummation procedure involves a parameter $\alpha$ to improve convergence [17,18], which is determined so as to reach optimal convergence of the resummed first- and second-order results. This here yields $\alpha=0.53$. For the estimation of error bars for the resummed result further orders of perturbation theory would be necessary, so one has to be cautious not to overestimate the precision of the results especially for larger $\overline{w}_R$. The result of the resummation is displayed as the middle curve in Fig. 1 and can be parametrized following [18] by

$$W(\overline{w}_R)=-2\epsilon_3\overline{w}_R+\frac{44}{15}\,\overline{w}_R^2\,\frac{1}{1+8.55\overline{w}_R}. \tag{27}$$

For $\overline{w}_R<0.5$ this parametrization reproduces the numerical results with a deviation less than 2% while for $\overline{w}_R>0.5$ the deviation increases up to 6%. Alternatively, one could use the Padé approximant to Eq. (19),

$$W(\overline{w}_R)=-2\epsilon_3\overline{w}_R+\frac{44}{15}\overline{w}_R^2$$
$$\times\frac{1}{1+15/44(3304/225+\pi^2 68/45)\overline{w}_R}, \tag{28}$$

which reproduces the $\overline{w}_R^2$ and $\overline{w}_R^3$ terms of $W(\overline{w}_R)$ and has the same qualitative behavior as Eq. (27) but lies about 15% below the resummed results. As expected, the result (27) shows again asymptotic free behavior at $d=3$ in the infrared limit with $\bar{w}_R=0$ being the only fixed point. For $d<3$ the nontrivial infrared stable fixed point emerges, whose fixed point value $w^*$ was given in Eq. (22) in the $\epsilon_3$ expansion.

The Borel resummation procedure also was carried through for the exponent functions in the interval $\bar{w}_R \in [0,1]$, leading to the parametrizations

$$\gamma_{m^2}(\bar{w}_R)=\frac{32}{675}\bar{w}_R^2\frac{1+0.257\bar{w}_R}{1+6.52\bar{w}_R}, \tag{29}$$

where in the resummation we choose $b=3.5$ and $\alpha=1.3$, and

$$\eta(\bar{w}_R)=\frac{2}{675}\bar{w}_R^2\frac{1+1.41\bar{w}_R}{1+2.19\bar{w}_R}, \tag{30}$$

with $b=2.0$ and $\alpha=3.53$.

### B. Integration of the flow equations for $d=3$ and $n=0$

Introducing the dimensionless scaling parameter $\lambda = l_0/l_R \in [0,1]$ and inserting Eq. (19) into Eq. (14) we can easily integrate the flow equation (14) via separation of variables. This gives for $n=0$ and $\epsilon_3=0$

$$\frac{1}{\bar{w}_R}-\left(\frac{826}{165}+\pi^2\frac{17}{33}\right)\ln\bar{w}_R+\mathcal{P}(\bar{w}_R)=s_3(\bar{w}_0)-\frac{44}{15}\ln\lambda, \tag{31}$$

where the polynomial $\mathcal{P}(\bar{w}_R)$ starting with terms of order $O(\bar{w}_R)$ contains all the (unknown) higher-order contributions to the Wilson function $W(\bar{w}_R)$. Setting $\lambda=1$ we find

$$s_3(\bar{w}_0)=\frac{1}{\bar{w}_0}-\left(\frac{826}{165}+\pi^2\frac{17}{33}\right)\ln\bar{w}_0+\mathcal{P}(\bar{w}_0) \tag{32}$$

for the integration constant $s_3(\bar{w}_0)$, which depends on the nonuniversal starting value $\bar{w}_0=\bar{w}_R(\lambda=1)$ of the renormalized coupling. Note that $\bar{w}_0$ is not equal to the coupling constant $w_0$ of the bare model.

If we neglect all but the leading $1/\bar{w}_R$ term for small $\bar{w}_R$ on the right hand side of Eq. (31) and use the result (51) for the relation between $\lambda$ and $N$ we find the lowest-order flow equation (1). Inverting our full result (31) we get the $\lambda$ dependence of the coupling $\bar{w}_R(\lambda)$ in an expansion for small $\lambda \ll 1$ as

$$\bar{w}_R(\lambda)=-\frac{15}{44\ln\lambda}\left[1-\left(\frac{413}{242}+\pi^2\frac{85}{484}\right)\times\frac{\ln(-\ln\lambda)}{\ln\lambda}+O\left(\frac{1}{\ln\lambda}\right)\right]. \tag{33}$$

The first two terms of the flow calculated in Eq. (33) are entirely universal, while the next term of order $O(\ln^{-2}\lambda)$ contains contributions of the nonuniversal integration constant $s_3(\bar{w}_0)$. Alternatively we can also integrate the resummed form (27) of the Wilson function which leads to the implicit equation

$$\ln\lambda=-\frac{15}{44}\left(-8.55\ln\bar{w}_R+\frac{1}{\bar{w}_R}\right)+C(\bar{w}_0) \tag{34}$$

for $\bar{w}_R(\lambda)$, with another integration constant $C(\bar{w}_0)$ replacing $s_3(\bar{w}_0)$.

Introducing

$$F_\eta(x)=\int\frac{\eta}{W}\quad\text{and}\quad F_{\gamma_{m^2}}(x)=-\int\frac{\gamma_{m^2}}{W}, \tag{35}$$

we integrate the exponent flow equations (15) and (16) to get

$$Z_\phi(\bar{w}_R)=s_\phi(\bar{w}_0)\exp[F_\eta(\bar{w}_R)]$$

and

$$Z_{m^2}(\bar{w}_R)=s_{m^2}(\bar{w}_0)\exp[F_{\gamma_{m^2}}(\bar{w}_R)], \tag{36}$$

with the nonuniversal integration constants

$$s_\phi(\bar{w}_0)=Z_\phi(\bar{w}_0)\exp[-F_\eta(\bar{w}_0)]$$

and

$$s_{m^2}(\bar{w}_0)=Z_{m^2}(\bar{w}_0)\exp[-F_{\gamma_{m^2}}(\bar{w}_0)]. \tag{37}$$

In an expansion for small $\bar{w}_R$ we find for $F_\eta$ and $F_{\gamma_{m^2}}$

$$F_\eta(\bar{w}_R)=\frac{1}{990}\bar{w}_R+\frac{1}{2}\left(\frac{997}{245025}+\frac{17\pi^2}{32670}\right)\bar{w}_R^2+O(\bar{w}_R^3), \tag{38}$$

$$F_{\gamma_{m^2}}(\bar{w}_R)=\frac{-8}{495}\bar{w}_R+\frac{1}{2}\left(\frac{4376}{245025}-\frac{449\pi^2}{81675}\right)\bar{w}_R^2+O(\bar{w}_R^3). \tag{39}$$

Similarly, we can calculate $F_\eta$ and $F_{\gamma_{m^2}}$ from Eq. (35) using the resummed results (27), (29), and (30) for $W(\bar{w}_R)$, $\gamma_{m^2}(\bar{w}_R)$, and $\eta(\bar{w}_R)$.

## IV. SCALING FUNCTIONS FOR POLYMER OBSERVABLES

For the mapping between renormalized and bare chain length in our polymer theory with the Hamiltonian (2) we define the $Z$ factor $Z_N$ via

$$l_0^2N=Z_Nl_R^2N_R. \tag{40}$$

With the mapping on the $\phi^6$ field theory we find the relation

$$Z_N=Z_{m^2}^{-1}. \tag{41}$$

Duplantier [3,4] has calculated the partition function, the squared end-to-end-vector, the radius of gyration, and the specific heat of a single chain up to first-order perturbation theory in the two-point and the three-point coupling. We here cite the renormalized results for vanishing effective two-point coupling in our nomenclature. For the squared end-to-end vector $R_e^2$, the radius of gyration $R_g^2$, and the universal ratio $R_e^2/R_g^2$ the results are

$$R_e^2=6Nl_0^2Z_N^{-1}\left(1-\frac{2}{15}\overline{w}_R+O(\overline{w}_R^2)\right)$$
$$=6\tilde{R}_0^2\left(1-\frac{74}{495}\overline{w}_R+O(\overline{w}_R^2)\right), \quad (42)$$

$$R_g^2=Nl_0^2Z_N^{-1}\left(1-\frac{13}{120}\overline{w}_R+O(\overline{w}_R^2)\right)$$
$$=\tilde{R}_0^2\left(1-\frac{493}{3960}\overline{w}_R+O(\overline{w}_R^2)\right), \quad (43)$$

$$R_e^2/R_g^2=6\left(1-\frac{1}{40}\overline{w}_R+O(\overline{w}_R^2)\right), \quad (44)$$

where the parameter $\tilde{R}_0^2=Nl_0^2 s_{m^2}(\overline{w}_0)$ contains the leading Gaussian chain length dependence $\sim N$ and the nonuniversal constant $l_0^2 s_{m^2}(\overline{w}_0)$. For the partition function $Z(N)$ at the tricritical point the result reads

$$e^{-\mu_S^* N}Z(N)=\frac{Z_\phi}{Z_N}\left(1-\frac{2}{15}\overline{w}_R+O(\overline{w}_R^2)\right)$$
$$=s_{m^2}s_\phi\left(1-\frac{49}{330}\overline{w}_R+O(\overline{w}_R^2)\right), \quad (45)$$

with the nonuniversal critical chemical segment potential $\mu_S^*$. Assuming that the strength of the two-point interaction is proportional to the deviation from the $\Theta$-point temperature $u_e \sim T-T_\Theta$ and that the three-point coupling is not temperature dependent one finds for the internal energy $E$ and for the specific heat $c_v$

$$E\sim\langle N_c^{(2)}\rangle \quad \text{and} \quad c_V\sim \text{const}\times\langle N_c^{(2)}\rangle+\text{const}\times\langle (N_c^{(2)})^2\rangle_c, \quad (46)$$

where $N_c^{(2)}$ is the number of two-point contacts of a given configuration. The mean value and second cumulant in Eqs. (46) can be calculated by taking the derivative of the partition function with respect to $u_e$. The results for $u_e=0$ are [12]

$$\langle N_c^{(2)}\rangle=-\mu_S^{(1)}(\overline{w}_0)N-2C_1(\overline{w}_0)N^{1/2}\overline{w}_R^{4/11}+\cdots, \quad (47)$$

$$\langle (N_c^{(2)})^2\rangle_c=NC_1^2(\overline{w}_0)\left(\frac{5\pi}{2}\overline{w}_R^{-3/11}+C_2(\overline{w}_0)-7.761\overline{w}_R^{8/11}+\cdots\right), \quad (48)$$

with the nonuniversal constants $\mu_s^{(1)}(\overline{w}_0)$, $C_1(\overline{w}_0)$, and $C_2(\overline{w}_0)$. In simulations one can measure directly the fluctuations of the number of two-point contacts (48), which has the advantage that the tricritical corrections affect already the leading behavior.

For a comparison of the field-theoretic results to Monte Carlo simulations we furthermore need the relation between the scaling parameter $\lambda=l_0/l_R$ and the bare chain length $N$. Equation (40) gives $\lambda=(Z_N N_R/N)^{(1/2)}$. Here we still have the freedom to choose the renormalized length scale $l_R$ at which we evaluate the renormalized perturbation theory. We choose it to be of the order of the smallest macroscopic length scale relevant to the observables we are interested in. For the global single-chain quantities, which we consider here, this is the radius of gyration. The choice

$$N_R=1 \quad (49)$$

leads to $l_R\approx R_g$ and gives

$$\lambda=Z_N^{1/2}N^{-1/2}\equiv Z_{m^2}^{-1/2}N^{-1/2}. \quad (50)$$

Using Eqs. (36), (39), and (33) we find for large $N$

$$\frac{1}{\ln\lambda}=\frac{-2}{\ln(Ns_{m^2})}+O(\ln^{-3}(Ns_{m^2}))=\frac{-2}{\ln(N)}+O(\ln^{-2}(N)). \quad (51)$$

## V. MONTE CARLO SIMULATIONS

There exists already a plethora of Monte Carlo simulations for various polymer models in the vicinity of the $\Theta$ point [5–8,19–24]. Some of these [7,8] claimed to have measured the logarithmic corrections to the Gaussian behavior as calculated by Duplantier [3,4]. Up to now the most elaborate and precise simulations are due to Grassberger and Hegger [5,6], who use a model of self-avoiding walks on a cubic lattice with an attractive nearest neighbor interaction between nonbonded nearest neighbors, simulating chain lengths up to $N=10\,000$ for single free chains. They found corrections to the Gaussian behavior that can be interpreted as logarithmic but with amplitudes up to 10 times larger than the universal amplitudes predicted by Duplantier. In [6] Grassberger also refuted claims by other authors [7,8,23] that the large corrections are due to lattice artifacts and are absent in off-lattice models or for lattices with higher coordination number. By simulating various models for longer chain lengths he found that the authors of [7,8,22–24] slightly underestimated the $\Theta$ temperature of their models, leading to smaller effective corrections. This shows a subtle point of the analysis of Monte Carlo simulations. To estimate the precise tricritical $\Theta$ temperature, which is a property of chains of infinite length, from simulations of chains of finite chain length one has to use the predictions of field theory for the extrapolation to infinite chain length. To get a reliable estimate one must be sure to be in a chain length regime where subleading nonuniversal corrections are negligible. This was apparently not the case in [7,8,22–24].

### A. Domb-Joyce-model

To check the universality of the large corrections found in [5,6] and to compare with the field-theoretic predictions we performed new Monte Carlo simulations on a Domb-Joyce-like model [25] of a weakly interacting nonreversal random walk (NRW) on a simple cubic lattice. As the basic ensemble we choose NRW's without immediate U turns instead of simple random walks since the algorithm showed a slightly better performance using the former. To model the two- and three-point interactions we weighted each walk of length $N$ (steps on the lattice) with a factor $W_N=(1-p)^{\kappa_2}(1-q)^{\kappa_3}$, where $p,q\in[-\infty,1]$ are interaction parameters. $\kappa_2$ or $\kappa_3$ counts the number of intersections of two or three strands of the chain, respectively. An $m$-fold intersection counts as $m(m-1)/2$ twofold intersections and as $m(m-1)(m-2)/6$ threefold intersections. The partition function of a chain of length $N$ is

$$Z(N)=\sum_{\{NRW\ config.\}} W_N=\sum_{\{NRW\ config.\}} (1-p)^{\kappa_2}(1-q)^{\kappa_3}. \quad (52)$$

For $p=q=0$ we get back the statistics of NRW's and for $p=1$ we get the statistics of self-avoiding walks. For a simple repulsive two-point-interaction $p\geqslant 0, q=0$ the model describes the full crossover between noninteracting and excluded volume polymers including the two branches of the crossover scaling functions predicted by field theory [26]. In the sequel we use a repulsive three-point interaction with $q=0.4$, counterbalanced by an attractive two-point interaction with $\ln(1-p)\in[0.05,0.105]$. For small $p$ this leads to excluded volume behavior, while for large $p$ a collapsed globular state occurs, separated from the former through the Θ transition. With the two interaction parameters $p$ and $q$ at hand it should be possible to cover the whole crossover from pure random walk behavior to tricritical behavior. With our choice of $p$ and $q$ we tried to achieve a starting value $\bar{w}_0$ of the renormalized coupling, which is small enough for the asymptotic predictions of field theory to be valid, but still large enough to produce measurable tricritical effects in the available range of chain length.

### B. Chain growth algorithm

To get an estimate for the partition function (52) and for other observables one can employ simple sampling, creating a NRW by adding a new segment at one end in a direction randomly chosen out of the five allowed directions. Recording the weight $W_N$ and other observables in each step one thereby collects data for all chain lengths between 1 and a maximal chain length $N_{\max}$. For nonvanishing values of the interaction parameters $p$ and $q$ the weight $W_N$ for a typical NRW is usually much smaller than the weight $W_N$ of a typical member of the ensemble we wish to sample. This problem of attrition is well known in the self-avoiding walk case $(p=1,\ q=0)$, where the number of sampled self-avoiding walks (SAW's) with $W_N=1$ decreases exponentially with the chain length, while all the other NRW's with self-intersections do have $W_N=0$ and do not contribute to the sample. This leads to large sample fluctuations for longer chain lengths $N\gtrsim 100$ since the sample is dominated by a few members of comparably large weight [27]. Several well-known methods such as enrichment [28] or biased Rosenbluth sampling [29] have been devised to circumvent this problem. Recently they were combined to the recursive pruned enriched Rosenbluth algorithm (PERM) [6], which up to now has been the most efficient algorithm for the simulation of long Θ polymers. In our simulations we used a slightly simpler pruned enriched simple sampling algorithm consisting mainly out of the recursively called routine STEP(N) described in Appendix B. The algorithm essentially performs a random walk in the space of chain lengths between the reflecting boundaries 0 and $N_{\max}$ [6]. The bunch of data accumulated between starting from 0 and reaching 0 again (called tour) is highly correlated, but different tours are uncorrelated. Sample quality therefore is measured in numbers of tours and given in Table I for the chosen parameter values of $p$.

TABLE I. Sample characteristics of the Θ-point simulations.

| $\ln(1-p)$ | Number of started tours | Number of tours that reached $N_{max}=10\,000$ |
|---|---|---|
| 0.105 | $10^7$ | 352960 |
| 0.1025 | $10^7$ | 370519 |
| 0.1015 | $10^7$ | 396965 |
| 0.1010 | $10^7$ | 373048 |
| 0.1 | $8\times 10^6$ | 298620 |
| 0.08 | $10^6$ | 35720 |
| 0.05 | $10^6$ | 33389 |

## VI. COMPARISON OF FIELD THEORY AND MONTE CARLO SIMULATIONS

In the sequel we exploit the results of Sec. IV for the scaling functions of polymer observables in two different ways. First, we expand the whole scaling function including $Z$ factors in the renormalized coupling $\bar{w}_R$ and insert consistently the expansions (33) of $\bar{w}_R$ in λ and (51) of λ in $N$. Thereby we get, besides the leading logarithmic corrections calculated by Duplantier [3,4], the universal subleading corrections proportional to $\ln(\ln N)/\ln^2 N$. As mentioned earlier this is the complete description of the universal behavior near the Gaussian fixed point available, since the next term of order $O(\ln^{-2} N)$ already contains contributions of the nonuniversal starting value $\bar{w}_0$ of the renormalized coupling. Second, we try to preserve as much as possible of the structure of the RG mapping using the resummed form (34) for the flow of the renormalized coupling $\bar{w}_R(\lambda)$ with some starting value $\bar{w}_0$ and the integrated $Z$ factors given by Eqs. (36) and (37) without expanding them in the scaling functions. This procedure contains the independent fit parameters $\bar{w}_0$, $s_\phi(\bar{w}_0)$, and $s_{m^2}(\bar{w}_0)$, because of the unknown $\bar{w}_0$ dependence of $Z_\phi(\bar{w}_0)$ and $Z_{m^2}(\bar{w}_0)$. We only know that the values $Z_\phi(\bar{w}_0)=Z_{m^2}(\bar{w}_0)=1$ are exact for $\bar{w}_0=0$. We start the analysis with the universal ratio $R_e^2/R_g^2$, where the nonuniversal parameters $s_\phi(\bar{w}_0)$ and $s_{m^2}(\bar{w}_0)$ drop out.

### A. Universal ratio $R_e^2/R_g^2$

The expansion of Eq. (44) with respect to $N$ leads to the universal prediction

$$\frac{R_e^2}{R_g^2}=6\left(1-\frac{3}{176}\frac{1}{\ln N}-\frac{3}{176}\times\left(\frac{413}{121}+\pi^2\frac{85}{242}\right)\frac{\ln(\ln N)}{\ln^2 N}+O(\ln^{-2} N)\right). \quad (53)$$

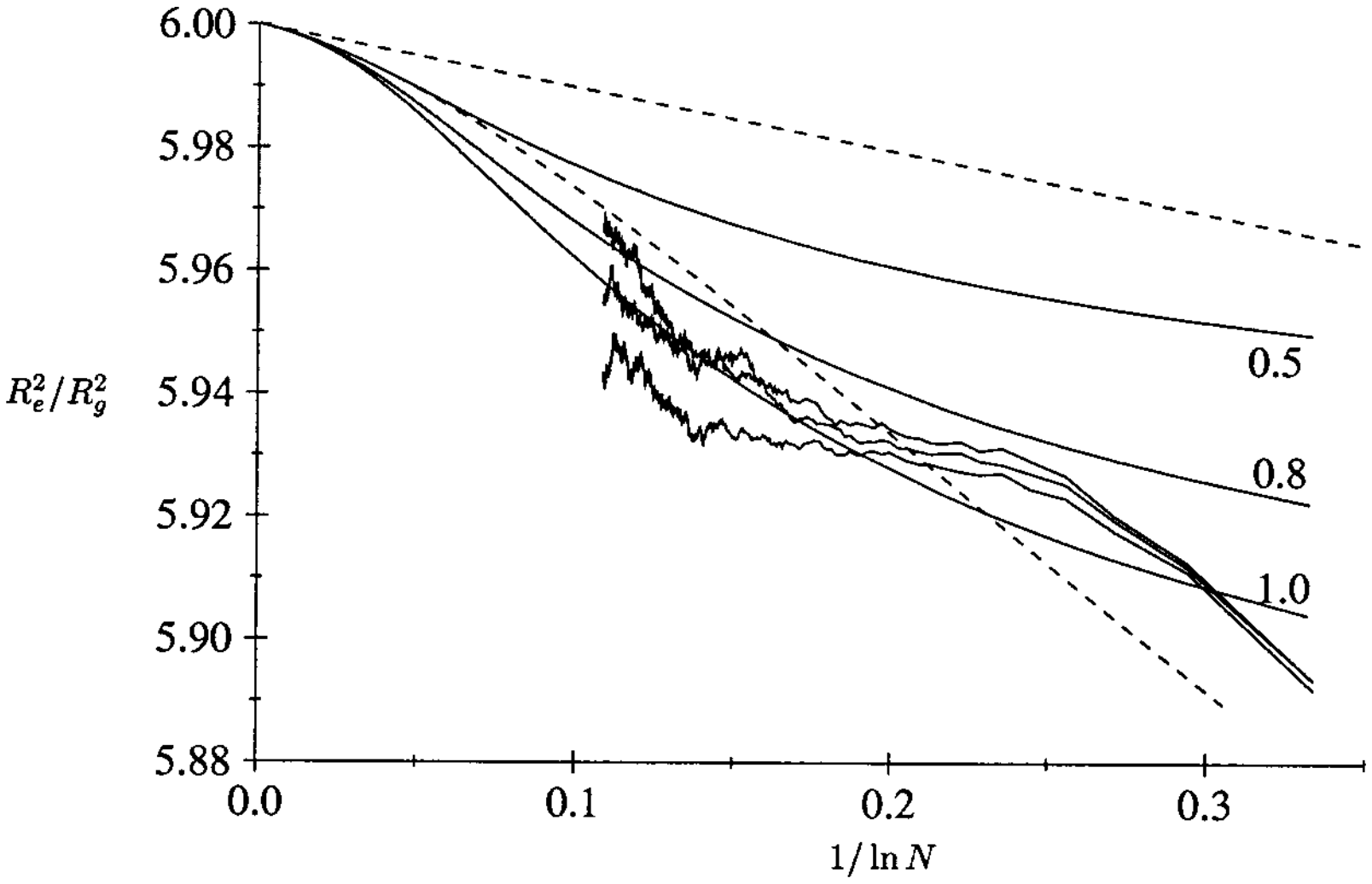

FIG. 2. Universal ratio $R_e^2/R_g^2$ plotted versus $1/\ln N$.

Figure 2 displays the simulation data for the parameter values $\ln(1-p)=0.1$, 0.1015, 0.105 (from top to bottom), which are close to the $\Theta$ point.The dashed and solid curves bundled on the left hand side give the theoretical results as discussed below. The approach to the fixed point value $R_e^2/R_g^2=6$ is from below as predicted by the universal result (53). The straight dashed line contains only the leading term of the universal result, while the curved dashed line includes also the subleading term. Note that the subleading term is larger than the leading one for all chain lengths reachable in simulations or experiments. Therefore, one cannot expect to measure the leading logarithmic correction. The full universal prediction of Eq. (53) lies not too far from the simulation data, but we regard this as an accidental coincidence due to the choice of the three-body parameter $q=0.4$ in our simulations. Preliminary results for smaller values of $q$ show smaller corrections, which are in fact closer to the leading term of Eq. (53). These findings preclude an explanation of the existing data only by the universal terms given in Eq. (53) and indicate that terms of order $O(\ln^{-2}N)$ are not negligible. The results of the full evaluation of the resummed theory are shown as solid curves for start couplings $\bar{w}_0=0.5$, 0.8, 1.0. The value $\bar{w}_0=1.0$ fits the data for $\ln(1-p)=0.1015$ quite well for $N\gtrsim 200$ within the rather large scatter of the data.

Note that there is a small hump in $R_e^2/R_g^2$ for $N\lesssim 150$ which we believe is due to chain end effects, not predicted by the tricritical theory. For most of the simulations presented in [5,6] this hump is much more pronounced and extents to longer chain length, leading even to a nonmonotonic behavior of $R_e^2/R_g^2$. This indicates that in those models with fairly large effective interactions the tricritical behavior may be blurred by strong short-chain effects.

### B. Radius of gyration

The radius of gyration of a discrete chain of $N$ segments (and therefore $N+1$ beads) is defined as

$$R_g^2=\frac{1}{N+1}\left\langle \sum_{i=0}^{N}[\mathbf{r}(i)-\mathbf{G}]^2\right\rangle, \tag{54}$$

with $\mathbf{G}=[1/(N+1)]\Sigma_{i=0}^{N}\mathbf{r}(i)$ being the center of mass. For the pure NRW on a cubic lattice one easily finds

$$R_g^2=\frac{32N^3+56N^2+44N-5+5^{(1-N)}}{128(N+1)^2}=\frac{1}{4}N[1+O(1/N)] \tag{55}$$

in units of the lattice constant, showing Gaussian behavior $\sim N$ with $l_0^2=1/4$ and corrections of order $O(1/N)$. Expanding the tricritical result (43) we find

$$\frac{R_g^2}{N}=l_0^2 s_{m^2}\left[1-\frac{493}{5808}\frac{1}{\ln N}-\frac{493}{5808}\left(\frac{413}{121}+\pi^2\frac{85}{242}\right)\times\frac{\ln(\ln N)}{\ln^2 N}+O(\ln^{-2}N)\right]. \tag{56}$$

For the microscopic length scale we choose $l_0^2=1/4$ as in the NRW case. Figure 3 displays the Monte Carlo results for the radius of gyration together with the theoretical predictions. The solid curves bundled on the right hand side are from top to bottom the simulation data for the parameter values $\ln(1-p)=0.1$, 0.1010, 0.1015, 0.1025, 0.105 of the attractive two-point interaction. We displayed data for chain lengths between 10 and 10 000 versus $1/\ln N$. Clearly one sees the beginning collapse of the polymer for $\ln(1-p)=0.105$. The two dashed curves bundled on the left hand side show the universal predictions of Eq. (56), where the parameter $s_{m^2}$, which induces only a multiplicative vertical shift, was set equal to 1. The straight upper curve contains only the leading logarithmic correction and the lower curve includes also the universal subleading term. Again we find that the subleading term is of the same order of magnitude as the leading correction for all chain lengths available in present simulations, which means that the true asymptotic behavior cannot be observed. Further one can see that the choice $s_{m^2}=0.984$ would provide a good fit of the leading logarithmic correction to the data for $\ln(1-p)=0.1025$ for chain lengths $N\gtrsim 150$, but since the subleading corrections dominate, such a fit has no meaning. Thereby using only the leading correction leads to an underestimation of the $\Theta$ temperature. The remaining three solid curves in Fig. 3 display the result of the evaluation of Eq. (45) using the integrated $Z$ factor $Z_{m^2}$ as given in Eq. (36) and the resummed coupling (34) for the starting values $\overline{w}_0=0.5$, 0.8, 1.0. With these values Eq. (37) yields $s_{m^2}/Z_{m^2}(\overline{w}_0)=1.01$, 1.017, 1.022, respectively, and we choose $Z_{m^2}(\overline{w}_0)=1$. One can see that by changing the value of $\overline{w}_0$ a wide range of slopes for the effective corrections can be covered. For the radius of gyration plotted in Fig. 3 the value $\overline{w}_0=1.0$ consistent with the value chosen for $R_e^2/R_g^2$ gives a good fit for $\ln(1-p)=0.1010$, which is our estimate for the attractive two-point coupling at the Θ point. Since we used two fit parameters $\overline{w}_0$ and $Z_{m^2}(\overline{w}_0)$ to obtain this estimate, it is necessary to use data for additional observables as a check.

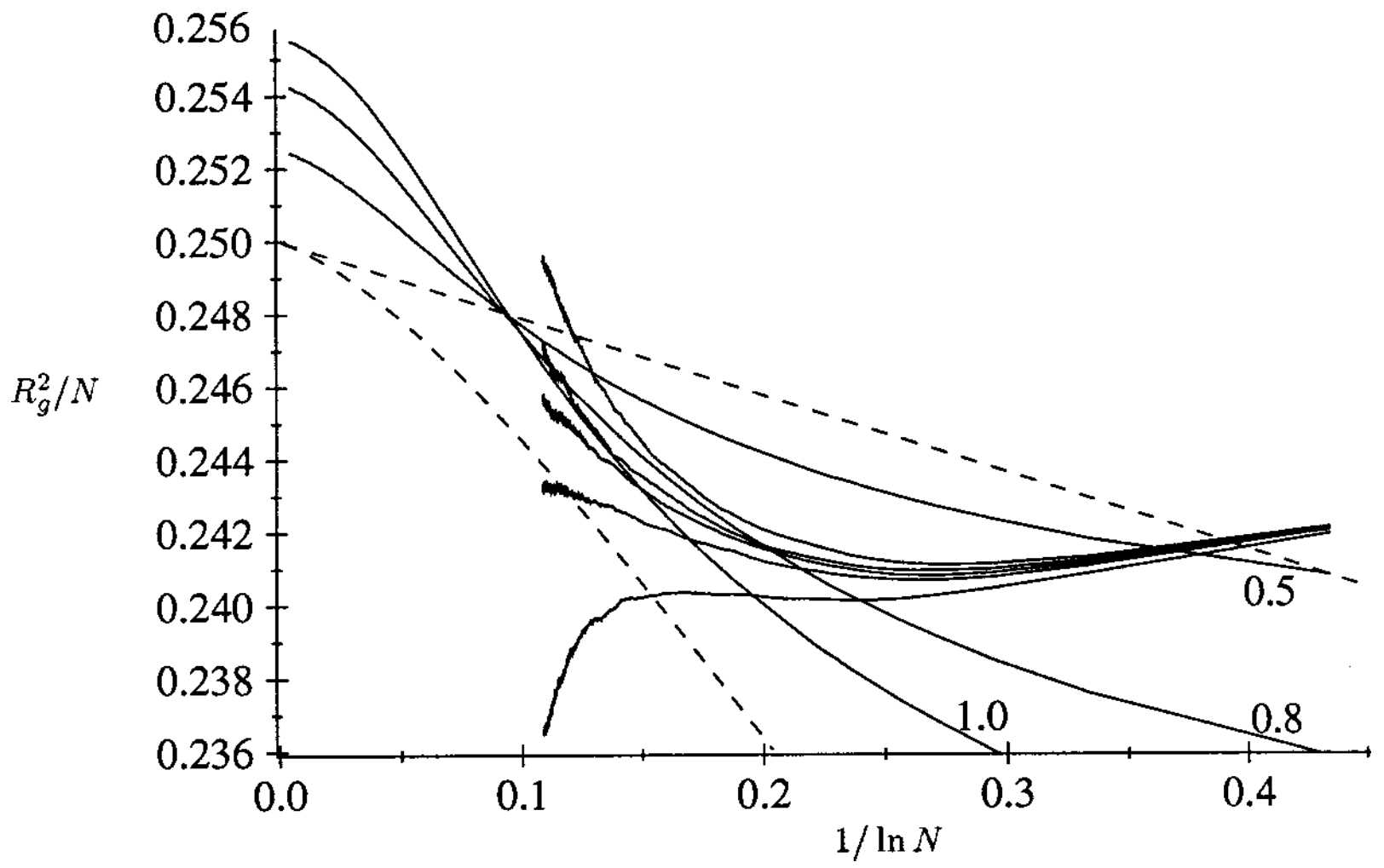

FIG. 3. Swelling factor $R_g^2/N$ for the radius of gyration plotted versus $1/\ln N$. $R_g$ is measured in units of the lattice constant.

### C. Squared end-to-end vector

For the squared end-to-end vector of the pure NRW we find

$$R_e^2=\langle[\mathbf{r}(N)-\mathbf{r}(0)]^2\rangle$$
$$=\frac{3}{2}N-\frac{5}{8}(1-5^{-N})=\frac{3}{2}N[1+O(1/N)], \qquad (57)$$

which shows again Gaussian behavior with $O(1/N)$ corrections. Expanding Eq. (42) we find the tricritical result

$$\frac{R_e^2}{N}=6l_0^2 s_{m^2}\Bigg[1-\frac{37}{363}\frac{1}{\ln N}-\frac{37}{363}\left(\frac{413}{121}+\pi^2\frac{85}{242}\right)\times\frac{\ln(\ln N)}{\ln^2 N}+O(\ln^{-2}N)\Bigg] \qquad (58)$$

for the swelling factor $R_e^2/N$. Figure 4 displays the simulation data and theoretical results for $R_e^2/N$ using the same parameters as in Fig. 3. The overall picture is very similar to that of Fig. 3. Again a fit using only the leading asymptotic correction of Eq. (57) would underestimate the Θ temperature. Our choice of $\overline{w}_0=1.0$ and $Z_{m^2}(\overline{w}_0)=1$ gives a quite reasonable fit for $\ln(1-p)=0.1010$, confirming our results for the radius of gyration.

### D. Partition function

To get rid of the unknown critical chemical segment potential $\mu_S^*$ in Eq. (45) we form the ratio $Z(N)^2/Z(2N)$. Expansion in $N$ gives

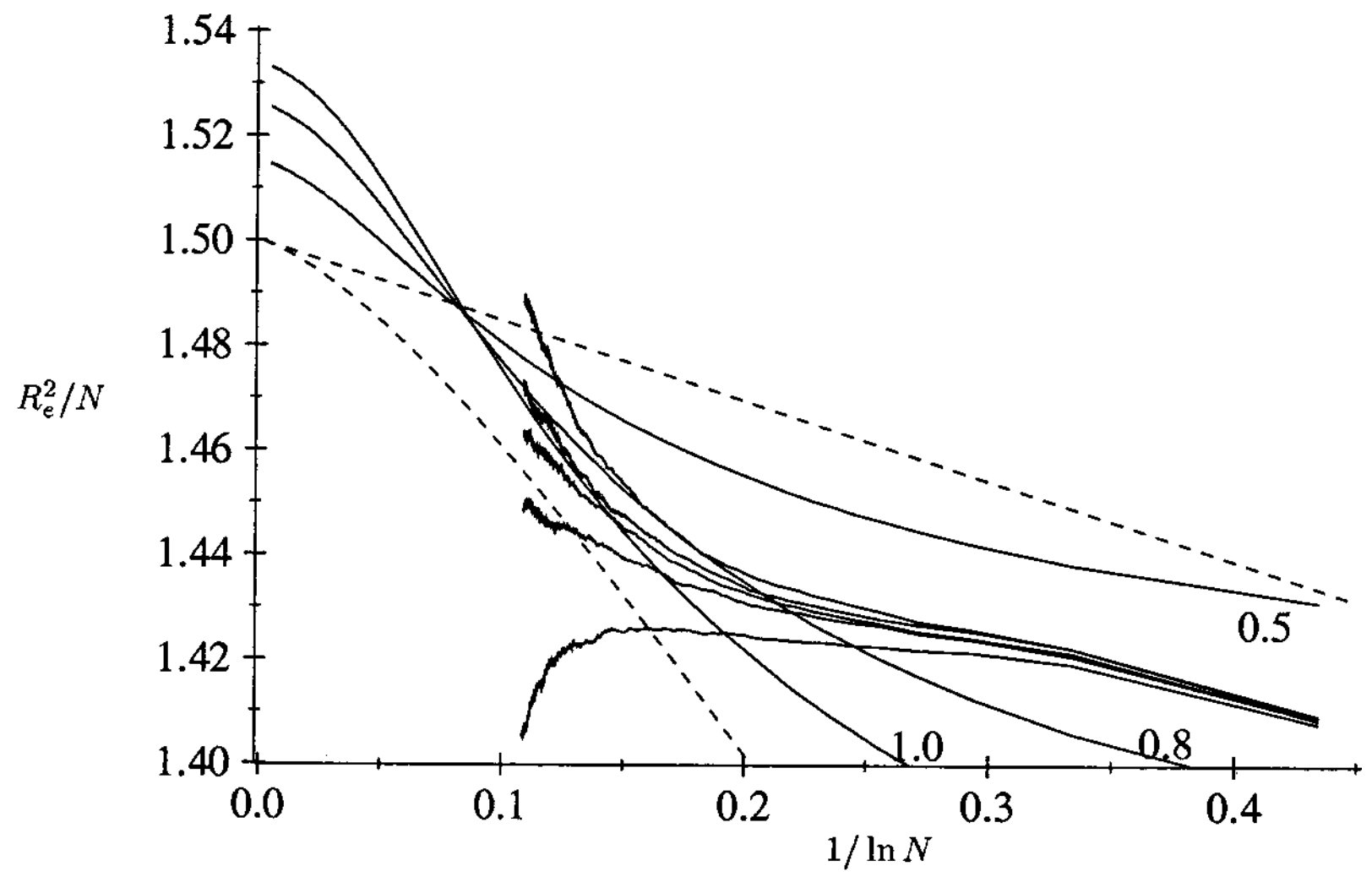

FIG. 4. Swelling factor $R_e^2/N$ for the end-to-end vector plotted versus $1/\ln N$.

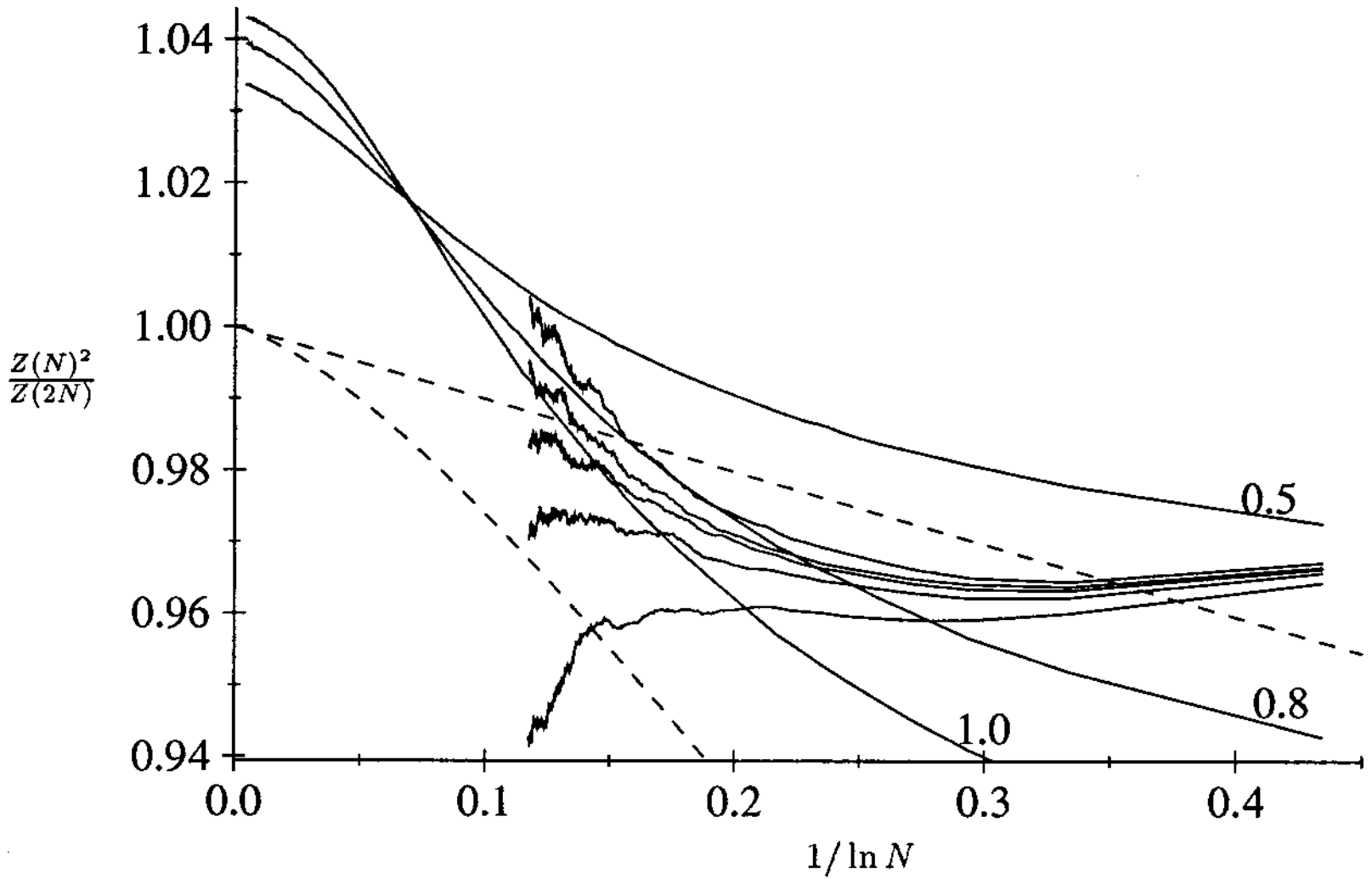

FIG. 5. Quotient $Z(N)^2/Z(2N)$ of the partition function plotted against $1/\ln N$.

$$\frac{Z(N)^2}{Z(2N)}=s_{m^2}s_\phi\left[1-\frac{49}{484}\frac{1}{\ln N}-\frac{49}{484}\left(\frac{413}{121}+\pi^2\frac{85}{242}\right)\times\frac{\ln(\ln N)}{\ln^2 N}+O(\ln^{-2}N)\right]. \tag{59}$$

Figure 5 displays the simulation data and the theoretical predictions using the same parameters as in Fig. 3. Concerning the universal predictions of Eq. (59) the situation is the same as for the radii. Our results for the unexpanded scaling functions (45) using the fit parameters $Z_{m^2}(\overline{w}_0)=1$ as before and $Z_\phi(\overline{w}_0)=1.025$ are shown as the solid lines labeled with their start couplings $\overline{w}_0=0.5$, 0.8, 1.0. A good fit of the data for $\ln(1-p)=0.1010$ for chain lengths $N\gtrsim 350$ can be achieved with $\overline{w}_0=0.95$, in fair agreement with the results for the radii. These findings differ from the results in [5,6], where the corrections for the partition sum were considerably smaller than those for the radii. One may speculate that this difference is due to terms of order $\overline{w}_R^2$ in the scaling functions, which become more important for the larger start coupling $\overline{w}_0$ that is necessary to explain the large corrections found in [5,6].

### E. Fluctuation of the number of two point contacts

In Fig. 6 we display the simulation data for $\langle (N_c^{(2)})^2\rangle_c/N$ plotted versus $\ln N$ for the parameter values $\ln(1-p)=0.105$, 0.1025, 0.1015, 0.1, 0.08, 0.05 (from top to bottom). For the values 0.05 and 0.08 which are already in the excluded volume regime one can see the expected saturation of $\langle (N_c^{(2)})^2\rangle_c/N$ at a finite value. The data for the parameter values near the $\Theta$ point show a pronounced deviation from the asymptotic Gaussian chain behavior

$$\langle (N_c^{(2)})^2\rangle_c \sim N\ln N, \tag{60}$$

but are still far away from the asymptotic tricritical behavior

$$\langle (N_c^{(2)})^2\rangle_c \sim N\ln^{3/11}N, \tag{61}$$

since the subleading terms $C_2(\overline{w}_0)$ and $-7.761\overline{w}_R^{8/11}$ in Eq. (48) are of the same order of magnitude as the leading term. Using $C_1(\overline{w}_0)$ and $C_2(\overline{w}_0)$ as fit parameters one can fit Eq. (48) perfectly well to the data for $\ln(1-p)=0.1015$ but due to two free fit parameters such a fit is of no significance. We find, even more than for the partition function and the universal ratio $R_e^2/R_g^2$, that the renormalized starting coupling $\overline{w}_0$ corresponding to our simulations is too large to see the asymptotic behavior. Here this clearly is not only due to corrections in the flow functions but also due to the higher orders in the scaling function. E.g., for $\overline{w}_0 = 1.0$ we find $\overline{w}_R \simeq 0.25$ for $N = 10\,000$ so that the subleading term $-7.761\overline{w}_R^{8/11}$ in Eq. (48) is only a factor of 4 smaller than the leading term $(5\pi/2)\overline{w}_R^{-3/11}$.

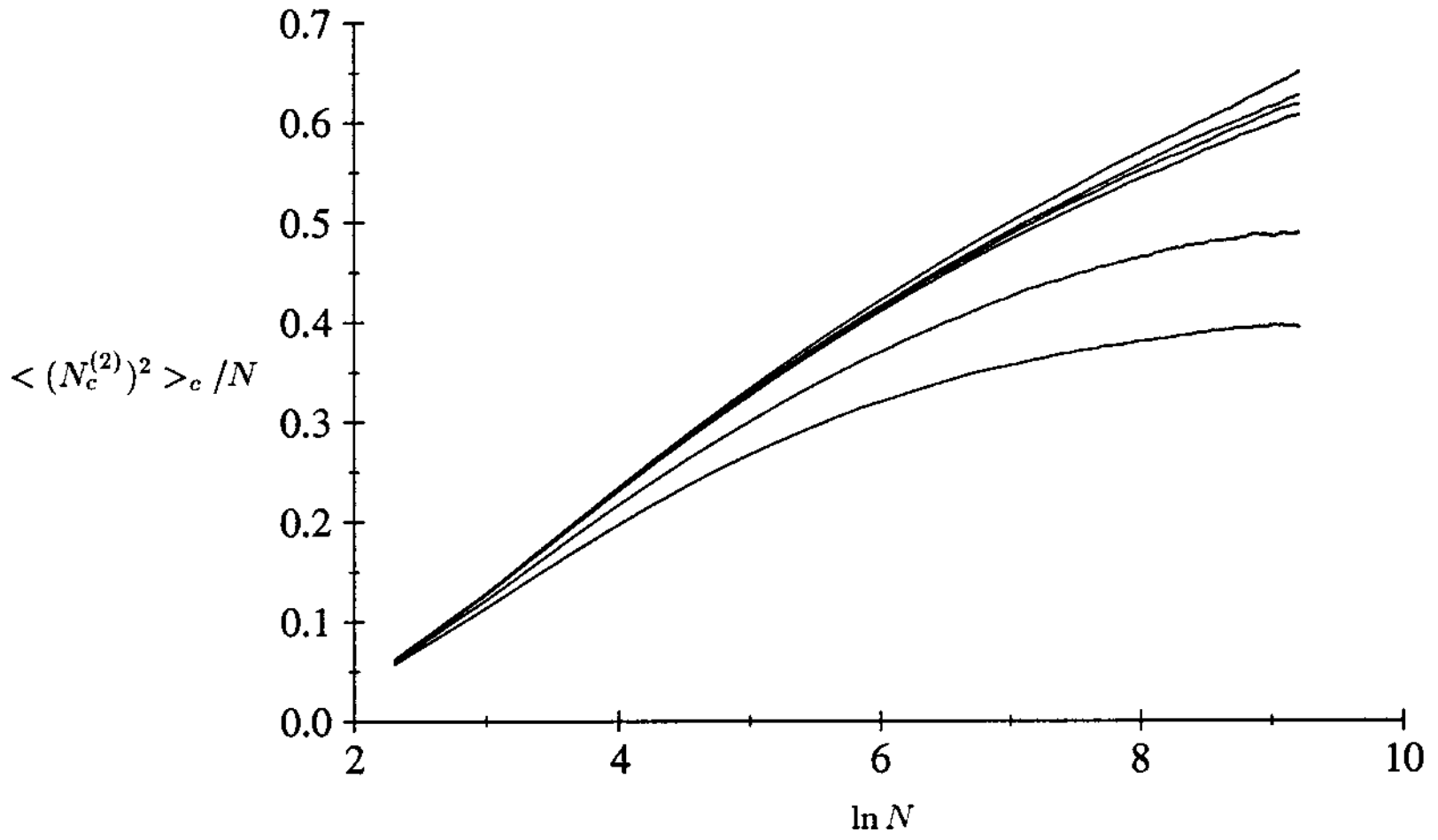

FIG. 6. Fluctuations of the number of two-point contacts plotted versus $\ln N$.

## VII. CONCLUSION AND OUTLOOK

The central question of this work concerns the interpretation of the strong quasilogarithmic corrections to Gaussian behavior, which are found in simulations of Θ-point polymers in three dimensions [5,6]. Can these corrections be explained within the tricritical approach? If not, we should seriously consider the possibility that the Θ transition in the standard polymer models is not in the universality class of zero-component $\phi^6$ theory. There is evidence for other scenarios in more complicated models [9–11].

To decide this question we pursued two different lines of approach. On the analytical side we calculated the subleading corrections of type $\ln(\ln N)/\ln^2 N$ within $\phi^6$ theory. We found that for all chain lengths available in present day simulations these corrections, which are still universal, are of the same order of magnitude as the leading $1/\ln N$ term. Therefore the previous data analysis, which employed only the leading corrections, is bound to fail. Furthermore, we evaluated the theory beyond the universal asymptotic region by combining known first-order results for the scaling functions with a Borel-resummed form of the RG mapping. This introduces some nonuniversal parameters like the starting value $\overline{w}_0$ of the renormalized coupling.

On the numerical side we performed new Monte Carlo simulations on a Domb-Joyce-like model of weakly interacting NRW's, including a repulsive three-body interaction $q$, compensated by a weakly attractive two-body interaction $p$. This model shows quasilogarithmic corrections which are much weaker than those found previously for models with built-in strict self-avoidance. We were able to consistently fit our theory to our data for $R_e^2$, $R_g^2$, and $Z$, for chain lengths $350 \lesssim N \lesssim 10^4$. For our choice $q = 0.4$ the Θ-point value of the two-body coupling is found as $\ln(1-p) = 0.1010(5)$. Clearly the chain length $N = 10^4$ reached in our simulations is not large enough to eliminate all doubts on the applicability of the tricritical theory. The problem is a delicate one, since we can determine the Θ point only by using the theoretical results. Still we feel that the overall agreement between theory and data is quite convincing.

Even if we take into account the subleading universal corrections or the full RG flow, our analytical results do not match the data of [5,6]. For $R_e^2$, for instance, these data show quasilogarithmic corrections which are nearly twice as strong as those found in our simulations. In the original publication the failure of the tricritical theory was interpreted as showing the existence of strong effective many-body forces, i.e., of interactions irrelevant in the sense of the renormalization group. Such terms certainly will play a role for shorter chains, $n \lesssim 500$, where the data for $R_e^2/R_g^2$ displayed in [5,6] show nonmonotonic behavior. However, the influence of such terms dies out roughly like $1/N$, and to us it seems unlikely that they seriously can distort the data for $N \gtrsim 1000$. We rather want to point to another source of the discrepancy. The strict self-avoidance, built into the previously used models, is likely to result in a large starting value $\overline{w}_0$ of the renormalized three-body coupling, and for the chain length reached also the running coupling constant $\overline{w}_R = \overline{w}_R(N, \overline{w}_0)$ will not be particularly small. This has two consequences. First, even our resummed RG mapping presumably is not very precise for larger $\overline{w}_R$. It, however, is known that a very good representation of the RG flow is essential for quantitative agreement between theory and data [12]. For instance, the explanation of the corrections to the Gaussian behavior at the upper critical dimension $d_c = 4$ of the excluded volume problem [30] employed a most precise RG flow, based on a five-loop calculation. The second consequence of a large value $\overline{w}_R$ may be even more serious. The scaling functions are known only to first order in $\overline{w}_R$, and for large $\overline{w}_R$ the higher-order terms will come into play. For the third virial coefficient, which is the only quantity where the term of order $\overline{w}_R^2$ is known [12], this term comes with a very large coefficient. This suggests that the expansions of the scaling functions are not particularly well behaved. All together this shows that higher-order calculations are needed for a quantitative explanation of data for models with strict self-avoidance. We believe that this finding is not restricted to the zero-component limit relevant for polymer physics, but applies to tricritical $\phi^6$ theory generally. To explain quantitatively tricritical effects in real experiments—the λ transition in $^3$He-$^4$He mixtures is here the most promising candidate—one will inevitably need such higher-order calculations.

Our results show that there is no chance to observe the leading logarithmic corrections at the tricritical point. Even for chains of length $N = 10^6$ we find from Eqs. (33) and (51) $(s_{m^2} \simeq 1)$ that the subleading term $\ln(\ln N)/\ln^2 N$ is of the same size as the leading term, which indicates that also the nonuniversal terms of flow are needed. The best way to further test the tricritical theory may be to pursue the path shown in the present paper. Choosing an even smaller three-body coupling $q$ we are sure to reach a regime where both the low-order expressions for the scaling functions and our approximate resummed RG flow are adequate. For the available chain lengths we then will not reach the universal regime and to test the theory we should measure as many quantities as possible ($R_e^2$, $R_g^2$, $Z$, second and third virial coefficients) and try to fit the data to the crossover scaling functions by adjusting the few nonuniversal parameters. The problem with this approach is that the effects will be very small and very good statistics is needed—much better than that of the present simulations.

## ACKNOWLEDGMENT

This work was supported by the Deutsche Forschungsgemeinschaft SFB 237 Unordnung und grosse Fluktuationen.

## APPENDIX A: CALCULATION OF THE $Z$ FACTORS

The actual calculation of the $Z$ factors uses the $2M$-point vertex functions $\Gamma^{(2M)}$ obtained from the connected $2M$-point Green's functions via Legendre transformation [31]. These renormalize according to

$$\Gamma^{(2M)}(\mathbf{k}_1,\ldots,\mathbf{k}_{2M},m_0^2,w_0,l_0) = Z_\phi^{-M}\Gamma_R^{(2M)}(\mathbf{k}_1,\ldots,\mathbf{k}_{2M},m_R^2,w_R). \qquad \text{(A1)}$$

Note that in this appendix we use the renormalized coupling $w_R = 2^5\pi^2\overline{w}_R$, thereby keeping the geometric factor $1/(2^5\pi^2)$ explicit. The $Z$ factors of the multiplicative RG mapping (37)–(41) are chosen to absorb all logarithmic divergences in the perturbation expansion of the vertex functions $\Gamma^{(2)}$ and $\Gamma^{(6)}$. (Note that additional multiplicative renormalizations are necessary for $\Gamma^{(4)}$ in the case of nonvanishing $\phi^4$ coupling $u_e$ [3,12,13], which is not considered here. Further divergent vertex functions show up if one considers $\phi^2$ and $\phi^4$ insertions.) To regularize the momentum integrals of perturbation theory we use dimensional regularization by analytical continuation from $d<2$, which automatically includes the necessary additive renormalization of $\Gamma^{(2)}$. The logarithmic divergences at $d_c=3$ then show up as poles in $\epsilon_3 = 3-d$. As renormalization scheme we chose minimal subtraction of $\epsilon_3$ poles. Similar calculations have been performed earlier with a different more complicated regularization and renormalization scheme to calculate the $\epsilon_3$ expansion of the tricritical exponents [13,32]. The diagrammatic perturbation series including only divergent diagrams is

$\Gamma^{(2)} = m_R^2 + k^2 -$ 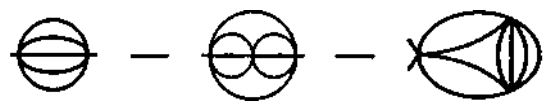 $+ O(w_R^4)$ (A2)

for $\Gamma^{(2)}$ up to order $w_R^3$ and

$\Gamma^{(6)} = -$ 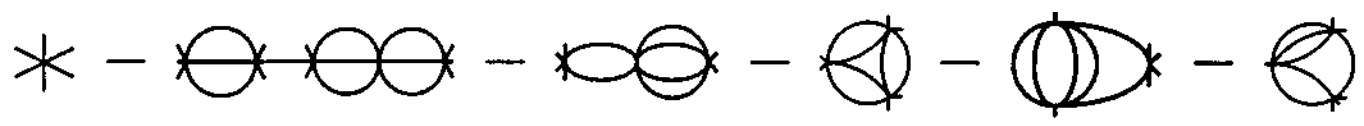 $+ O(w_R^4)$ (A3)

for $\Gamma^{(6)}$ up to order $w_R^3$. We omitted all diagrams containing tadpole subdiagrams, where a propagator begins and ends at the same vertex, since they do not contribute to the $Z$ factors. As part of the general BPHZ proof of renormalizability one finds that all divergences in $\Gamma^{(2)}$ and $\Gamma^{(6)}$ are removed by the $Z$ factors defined as

$$Z_\phi = 1 - \frac{\partial}{\partial k^2}\mathcal{K}\overline{R}\,\Gamma^{(2)}(\mathbf{k},m_R^2,w_R), \qquad \text{(A4)}$$

$$Z_2 = 1 - \frac{\partial}{\partial m_R^2}\mathcal{K}\overline{R}\,\Gamma^{(2)}(\mathbf{k},m_R^2,w_R), \qquad \text{(A5)}$$

$$Z_6 = 1 - \frac{1}{w_R}\mathcal{K}\overline{R}\,\Gamma^{(6)}(\mathbf{k}_i,m_R^2,w_R), \qquad \text{(A6)}$$

where the $\mathcal{K}$ operation picks out the pole part of a expression and the incomplete Bogoliubov $\overline{R}$ operation recursively subtracts the divergences of all divergent subdiagrams. For a detailed account on the procedure in the case of $\phi^4$ theory up to five loops we refer to [33,34]. To simplify the calculation of the momentum integrals we evaluated them in the massless case $m_R=0$, since this leads to a factorization in most of the Feynman integrals. Setting $m_R=0$ might lead to artificial IR singularities in the momentum integrals, which one has to avoid by properly chosen external momenta or to subtract by additional IR counterterms. For the calculation of Eq. (A5) one has to perform the derivative $\partial/\partial m_R^2$, which commutes with the $\mathcal{K}\overline{R}$ operation, before setting $m_R=0$. The differentiation of a propagator produces a $-\phi^2$ insertion:

$\frac{\partial}{\partial m_R^2}\frac{1}{p^2+m_R^2} = \frac{-1}{(p^2+m_R^2)^2} \hat{=} \frac{\partial}{\partial m_R^2}$  . (A7)

From Eq. (A2) we find

$\frac{\partial}{\partial m_R^2}\Gamma^{(2)} = 1 + 5$ 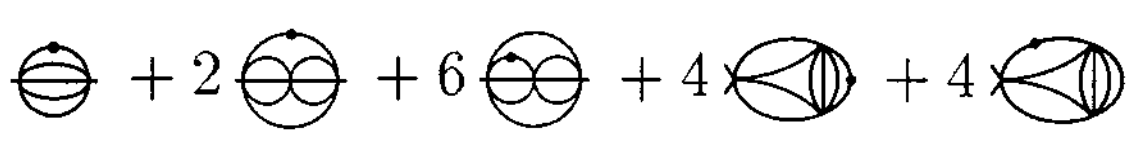 $+ O(w_R^4)$. (A8)

For $m_R=0$ the $\phi^2$ insertions cause IR singularities, which up to the given order can be removed by a suitable choice of the external momentum [33,34]. The UV counterterms corresponding to the diagrams of Eqs. (A2), (A3), and (A8) are

$-\mathcal{K}\bar{R}($ 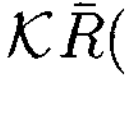 $) = -\mathcal{K}($ 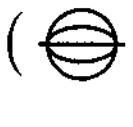 $) = \frac{w_R^2 k^2}{2^{10}\pi^4 6\epsilon_3}$ (A9)

$$-\mathcal{K}\bar{R}(\;) = -\mathcal{K}[\; - 2(\mathcal{K}[\;])\;] = \frac{w_R^3 k^2}{2^{15}\pi^6}\left(\frac{1}{9\epsilon_3^2} - \frac{2}{27\epsilon_3}\right) \tag{A10}$$

$$-\mathcal{K}\bar{R}(\;) = -\mathcal{K}(\;) = \frac{w_R^2}{2^{10}\pi^4 2\epsilon_3} \tag{A11}$$

$$-\mathcal{K}\bar{R}(\;) = -\mathcal{K}[\; - 2(\mathcal{K}[\;])\;] = \frac{w_R^3}{2^{15}\pi^6}\left(\frac{1}{3\epsilon_3^2} - \frac{2}{3\epsilon_3}\right) \tag{A12}$$

$$-\mathcal{K}\bar{R}(\;) = -\mathcal{K}[\; - (\mathcal{K}[\;])\;] = \frac{w_R^3}{2^{15}\pi^6}\left(\frac{1}{6\epsilon_3^2} - \frac{1}{\epsilon_3}\right) \tag{A13}$$

$$-\mathcal{K}\bar{R}(\;) = \frac{-w_R^3}{2^{15}\pi^4 6\epsilon_3} \tag{A14}$$

$$-\mathcal{K}\bar{R}(\;) = 0 \tag{A15}$$

$$-\mathcal{K}\bar{R}(\;) = -\mathcal{K}(\;) = \frac{-w_R^2}{2^5\pi^2\epsilon_3} \tag{A16}$$

$$-\mathcal{K}\bar{R}(\;) = \frac{-w_R^3}{2^{10}\pi^4\epsilon_3^2} \tag{A17}$$

$$-\mathcal{K}\bar{R}(\;) = 0 \tag{A18}$$

$$-\mathcal{K}\bar{R}(\;) = -\mathcal{K}(\;) = \frac{w_R^3}{2^{10}\pi^2 4\epsilon_3} \tag{A19}$$

$$-\mathcal{K}\bar{R}(\;) = -\mathcal{K}[\; - \underbrace{(\mathcal{K}\bar{R}[\;])}_{=0}\; - (4\mathcal{K}[\;])\underbrace{\;}_{=0}\;] = -\frac{w_R^3}{2^{10}\pi^4 2\epsilon_3} \tag{A20}$$

$$-\mathcal{K}\bar{R}(\;) = -\mathcal{K}[\; - (\mathcal{K}[\;])\;] = -\frac{w_R^3}{2^{10}\pi^4}\left(\frac{1}{2\epsilon_3^2} - \frac{1}{\epsilon_3}\right) \quad . \tag{A21}$$

The symmetry factors for each diagram can be calculated by multiplication of the factor $S_1$ for the case $n=1$ with the $n$-dependent factor $W_S$ that results from the summation of the inner spin indices of each diagram [33,34]. For our diagrams the symmetry factors are [13]

| diagram | | | | | | | | | |
|---|---|---|---|---|---|---|---|---|---|
| $S_1$ | $\frac{5}{3}$ | $\frac{5}{18}$ | $\frac{5}{16}$ | $\frac{15}{8}$ | $\frac{5}{8}$ | $5$ | $\frac{1}{120}$ | $\frac{1}{72}$ | $\frac{1}{192}$ |

(A22)

| dia. | 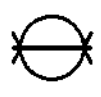 | 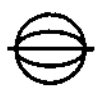 | 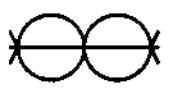 | 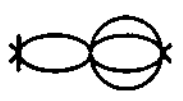 | 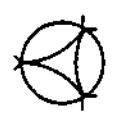 |
|---|---|---|---|---|---|
| $W_S$ | $\frac{3n+22}{25}$ | $\frac{n^2+6n+8}{15}$ | $\frac{27(3n^2+30n+92)}{15^3}$ | $\frac{9(n+4)(n^2+10n+64)}{15^3}$ | $\frac{n^3+34n^2+620n+2720}{15^3}$ |

(A23)

| 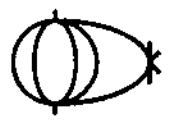 | 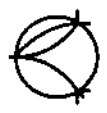 | 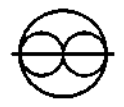 | 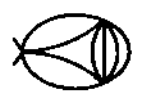 |
|---|---|---|---|
| $\frac{27(3n+22)(n+4)}{15^3}$ | $\frac{45n^2+702n+2628}{15^3}$ | $\frac{9(3n+22)(n+4)(n+2)}{15^3}$ | $\frac{3(3n+14)(n+4)^2(n+2)}{15^3}$ |

(A24)

The counterterms (A9)–(A21) together with the symmetry factors (A22)–(A24) inserted into Eqs. (A4)–(A6) finally lead to the $Z$ factors (11)–(13).

## APPENDIX B: PRUNED ENRICHED SIMPLE SAMPLING ALGORITHM

Here we briefly describe the main steps of the recursive routine STEP(N) of our pruned enriched simple sampling algorithm:

(i) On one end of a given walk of length $N-1$ we add a new segment by sampling one of the allowed directions, unless $N$ exceeds some value $N_{max}$, thereby creating a walk of length $N$. The coordinates of the new segment together with other necessary informations are stored in a hash table from which one can read out how often and from which directions a coordinate has been visited already.

(ii) The weight $W_N$ and other observables of interest are calculated and stored in cumulative lists.

(iii) The actual weight $W_N$ is compared to a guiding value $W_g(N)$, which may be the actual mean value of the sample or may be given by previous simulations or calculations. If the actual weight $W_N$ exceeds some multiple $r$ of the guiding value, we perform an enrichment step, which means that we set $W_N := W_N/2$ and call the routine STEP$(N+1)$ two times. If the actual weight $W_N$ is lower than some fraction $s$ of the guiding value, we perform a pruning step, which means that we set $W_N := 2W_N$ and call the routine STEP$(N+1)$ only with probability 1/2. If the actual weight $W_N$ is in the corridor $sW_g(N) \leq W_N \leq rW_g(N)$ we just call the routine STEP$(N+1)$ once.

(iv) Before we jump back from STEP(N) we clear all entries for the $N$th segment in the hash table but keep the entries in the cumulative lists.